\documentclass{emulateapj}
\usepackage{apjfonts}

\usepackage{natbib}
\usepackage{graphics}
\usepackage{multirow}
\usepackage{amsbsy}
\usepackage{mathrsfs}
\usepackage{footmisc}
\usepackage{hyperref}
\hypersetup{
  colorlinks = true,
  urlcolor = black,
  linkcolor=black,
  citecolor=black
}
\def\healpix{{\tt Healpix}}

\def\aap{Astr. \& Astroph.}

\def\apjl{Astrophys.\ J.\ Lett.}
\def\apjs{Astrophys.\ J.\ Suppl.}

\def\rwonsim{10,000}
\def\lmsbinonepval{0.1}

\def\lmsbinonetmin{84}
\def\lmsbinonetmax{87}
\def\lmsmaxpix{85.24}
\def\tmaxmean{120}

\def\tmaxpctabove{96}
\def\nzero{334}
\def\pctzero{3.3}
\def\nthousand{6273}
\def\pctthousand{63}

\def\lmsbinfourtmin{75}

\shorttitle{Cluster analysis of radio loops}
\shortauthors{R.~W.~Ogburn~IV}

\begin{document}

\title{Comment on cluster analysis of radio loops in CMB data}

\author{R.~W.~Ogburn~IV}
\affil{Department of Physics, Stanford University, Stanford, CA 94305, USA}
\affil{Kavli Institute for Particle Astrophysics and Cosmology, SLAC National Accelerator Laboratory, 2575 Sand Hill Rd, Menlo Park, CA 94025, USA}

\begin{abstract}
A recent article~\citep{LMS} looks for evidence in the WMAP internal linear combination
map (ILC) of unmodeled emission from the galactic radio loop known as Loop I.
The statistically strongest result comes from a cluster analysis that tests whether the
peak pixels within a $20^\circ$ annulus at Loop I are preferentially located
near the center line of the annulus.  From this cluster analysis the authors report a $p$-value of
0.018\% when considering the
four highest bins (75--87~$\mu$K).  I show that the reported statistical
significance has been overestimated.  First, the analysis
does not correctly select the hottest peaks in the simulated sky realizations;
second, it is sensitive to the map pixelization used, and in particular,
pixel size used is similar to the relevant clustering distance.
I have run $\rwonsim$ simulated sky realizations to reproduce the analysis
in \cite{LMS} and to calculate the effects of incorrect peak selection and
of pixelization.  Accounting for both of these effects, I find a 
$p$-value of $\sim 1\%$, both in the highest-bin test and in the four-bin test.
Finally, I note that even under the assumption that Loop I contributes significant power
to the ILC map, the observed clustering remains very unlikely.
Therefore, a result inconsistent with statistical
isotropy is not automatically strong evidence for a detection of Loop I.
I suggest additional tests that could clarify the degree to which the cluster analysis
supports a detection of Loop I in the CMB map.
\end{abstract}

%%%%%%%%%%%%%%%%%%%%%%%%%%%%%%%%%%%%%%%%%
\section{Introduction}

Several large-scale ``Loops'' associated with supernovae and other
processes within the galaxy have been detected in radio surveys~\citep{BHS}.
The most prominent is Loop I, joined by others in many
parts of the sky~\citep{W07}.
It has recently been reported~\citep[subsequently abbreviated as $LMS$]{LMS} that contamination associated with Loop I is also present
in the internal linear combination (ILC) map produced from multiple frequencies of the 9-year WMAP
data set~\citep{wmap9yr}.
The presence of such contamination is inferred from the mean temperature and skewness
within a small $2^\circ$ annulus at Loop I, with a 1--3\% probability of occurring
by chance.

A second type statistic is also used, based on the clustering of high-temperature peaks or ``hot spots'' around
the center line of Loop I.  The clustering analysis is similar to one that has previously
been used for study of possible foregrounds in the WMAP ILC~\citep{NDV}.
It is calculated that the hottest pixels in the ILC map cluster
tightly around the center line in a way that is duplicated in only
$\lmsbinonepval\%$ of simulations ($p$-value of $\lmsbinonepval\%$) for the highest-temperature bin,
or $p$-value of 0.018\% when combining four bins.

The structure of this note is as follows.
In Section~2 of this note I summarize the cluster analysis used in the $LMS$ paper.
In Section~3 I show that the published analysis has not correctly selected
the ``hot spots'' for clustering analysis of the 100,000 simulated sky realizations.
In Section~4 I recalculate the statistical significance of evidence for Loop I
contamination of the ILC, using $\rwonsim$ new simulated maps with modified binning.
In Section~5 I discuss the dependence of the result on the map
pixelization used and calculate the $p$-values from an additional set of simulations using
a finer map.  This analysis gives a $p$-value of $\sim 1\%$.
Finally, in Section~6 I discuss the likelihood of the observed values of the clustering
statistic with an interpretation as a detection of Loop I.

%-------------
   \begin{figure}
   \begin{center}
   \begin{tabular}{c}
   \includegraphics[width=0.96\columnwidth]{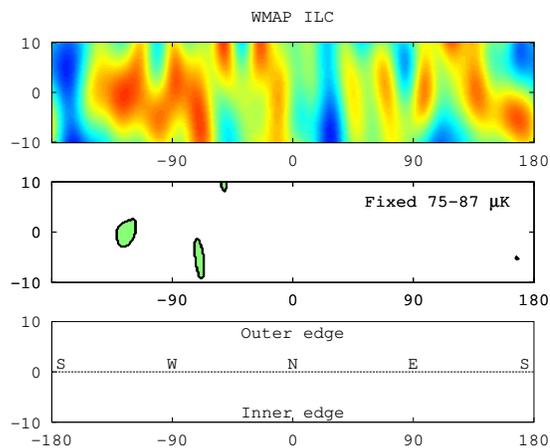}
   \end{tabular}
   \end{center}
   \caption[example]
   { \label{fig:strip_ilc}
     WMAP ILC map filtered at $\ell\le 20$ within the Loop I strip.  The data and region selection are the
     same as in Fig.~2 of $LMS$, but for convenience the annulus is flattened into a strip.
     The direction toward galactic north is in the center, and the outer edge of the strip is at the top.
     The color scale in the temperature map {\it (top panel)} is [$-130$,$+130~\mu\mathrm{K}$].
     The temperature range $\lmsbinfourtmin\le T<\lmsbinonetmax~\mu\mathrm{K}$ {\it (middle panel)}
     contains the four bins used for the clustering analysis in $LMS$.  By construction, these bins
     contain the hottest pixels within the Loop I region.
   }
   \end{figure}
%-------------

%-------------
   \begin{figure}[h]
   \begin{center}
   \begin{tabular}{c}
   \includegraphics[width=0.96\columnwidth]{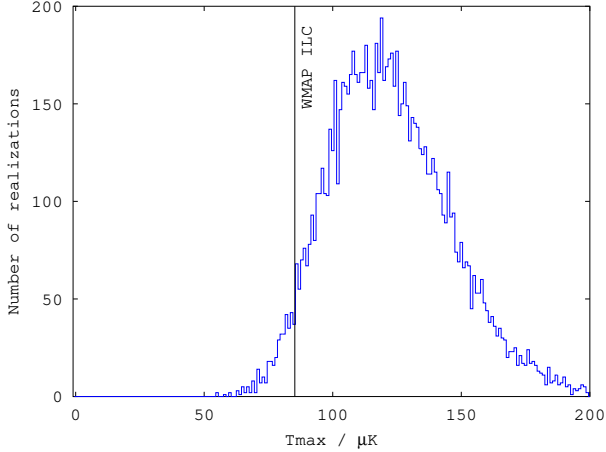}
   \end{tabular}
   \end{center}
   \caption[example]
   { \label{fig:tmax_hist}
     Maximum temperature in Loop I annulus in WMAP ILC and simulations
   }
   \end{figure}
%-------------

%%%%%%%%%%%%%%%%%%%%%%%%%%%%%%%%%%%%%%%%%
\section{Cluster analysis}

The cluster analysis in the $LMS$ paper uses the WMAP ILC map, filtered to include only
multipoles $\ell\le 20$ and downgraded to NSIDE=128~\citep{mertsch}.
An annulus is defined within $10^\circ$ of the center line of Loop I, according to the
parameters in~\citet{BHS}.
Fig.~\ref{fig:strip_ilc} shows the WMAP ILC map within this region after filtering.
I have chosen to flatten the loop into a strip for convenience in comparing the figures.
The top panel shows the CMB $T$ map, the middle panel shows the selected hot spots, and
the bottom panel indicates the orientation of this flattened strip relative to the sky in galactic coordinates.

The selection of this annulus can be described as follows.  Let $P$ be the set of \healpix\
pixels at NSIDE=128, and $\hat{n}_p$ be the unit vector to the center of any given pixel
$p \in P$.  Loop I is defined by the unit vector $\hat{n}_c$ of its center at
$l=329^\circ, b=+17.5^\circ$ and its radius $r=58^\circ$.  The annulus $A$ is then defined
as
\begin{equation}
\label{eq:hotspotsel}
A=\left\{p \in P ~|~ -10^\circ \le d(\hat{n}_p,\hat{n}_c)-r \le +10^\circ\right\}
\end{equation}
where $d(\hat{n}_1,\hat{n}_2)=\mathrm{Cos}^{-1}(\hat{n}_1 \cdot \hat{n}_2)$ is the
% arc length
great circle distance
between two points on the sphere.

For purposes of the cluster analysis the $LMS$ authors 
define bins in CMB temperature with $\Delta T=3~\mu\mathrm{K}$.
The pixels in these bins can be
defined as
\begin{equation}
A_i=\left\{p \in A ~|~ T_i \le T(\hat{n}_p) < T_i+\Delta T\right\}
\end{equation}
where $T(\hat{n})$ is the temperature in direction $\hat{n}$ in the filtered WMAP ILC map.
The highest four bins, i=1--4, are intended to select the ``hot spots'' within the
Loop region.  The middle panel of Fig.~\ref{fig:strip_ilc} shows the pixels within these four bins,
spanning $\lmsbinfourtmin$--$\lmsbinonetmax~\mu\mathrm{K}$.
The clustering statistic for a single bin is then defined by the average distance of selected
pixels from the Loop,
\begin{equation}
G_i = \frac{\sum_{p \in A_i} \left| d(\hat{n}_p,\hat{n}_c)-r) \right|}{N(p \in A_i)}.
\end{equation}
The $LMS$ paper uses $G_i$ for the clustering statistic defined relative to the filtered WMAP ILC
map, and $g_i$ for the same statistic defined relative to
a simulated sky realization prepared in the same way as the ILC map, with pixels selected
according to Eq.~\ref{eq:hotspotsel}.  In all cases the bins are defined using fixed bin lower
edges $T_i=75, 78, 81$, and $84~\mu\mathrm{K}$ for $i=1$--$4$.

%-------------
   \begin{figure}[t]
   \begin{center}
   \begin{tabular}{c}
   \includegraphics[width=0.96\columnwidth]{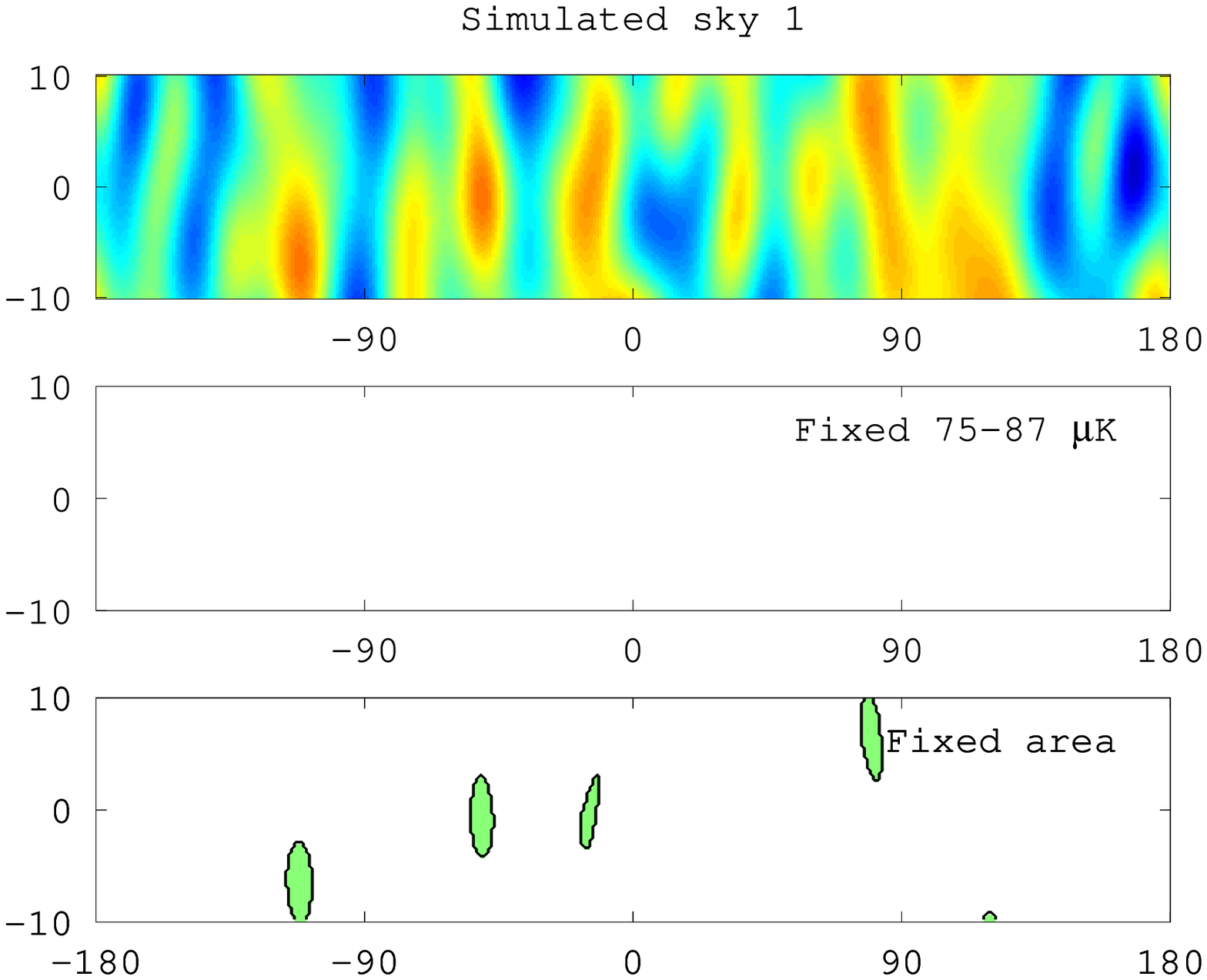}
   \end{tabular}
   \end{center}
   \caption[example]
   { \label{fig:strip_sim1}
      One simulated sky realization within the Loop I strip.  The color scale in the temperature map
      {\it (top panel)} is [$-130$,$+130~\mu\mathrm{K}$].  This realization has $T_\mathrm{max}$ lower than
      in the ILC map, and has no pixels within the range $\lmsbinfourtmin\le T<\lmsbinonetmax~\mu\mathrm{K}$ {\it (middle panel)}.
      About $\pctzero\%$ of the simulated sky maps fall into this category.  The analysis in $LMS$
      has counted them as less strongly clustered than in the ILC map.  The peaks can correctly be
      selected by taking the hottest pixels up to a fixed area {\it (bottom panel)}.
   }
   \end{figure}
%-------------

%-------------
   \begin{figure}
   \begin{center}
   \begin{tabular}{c}
   \includegraphics[width=0.96\columnwidth]{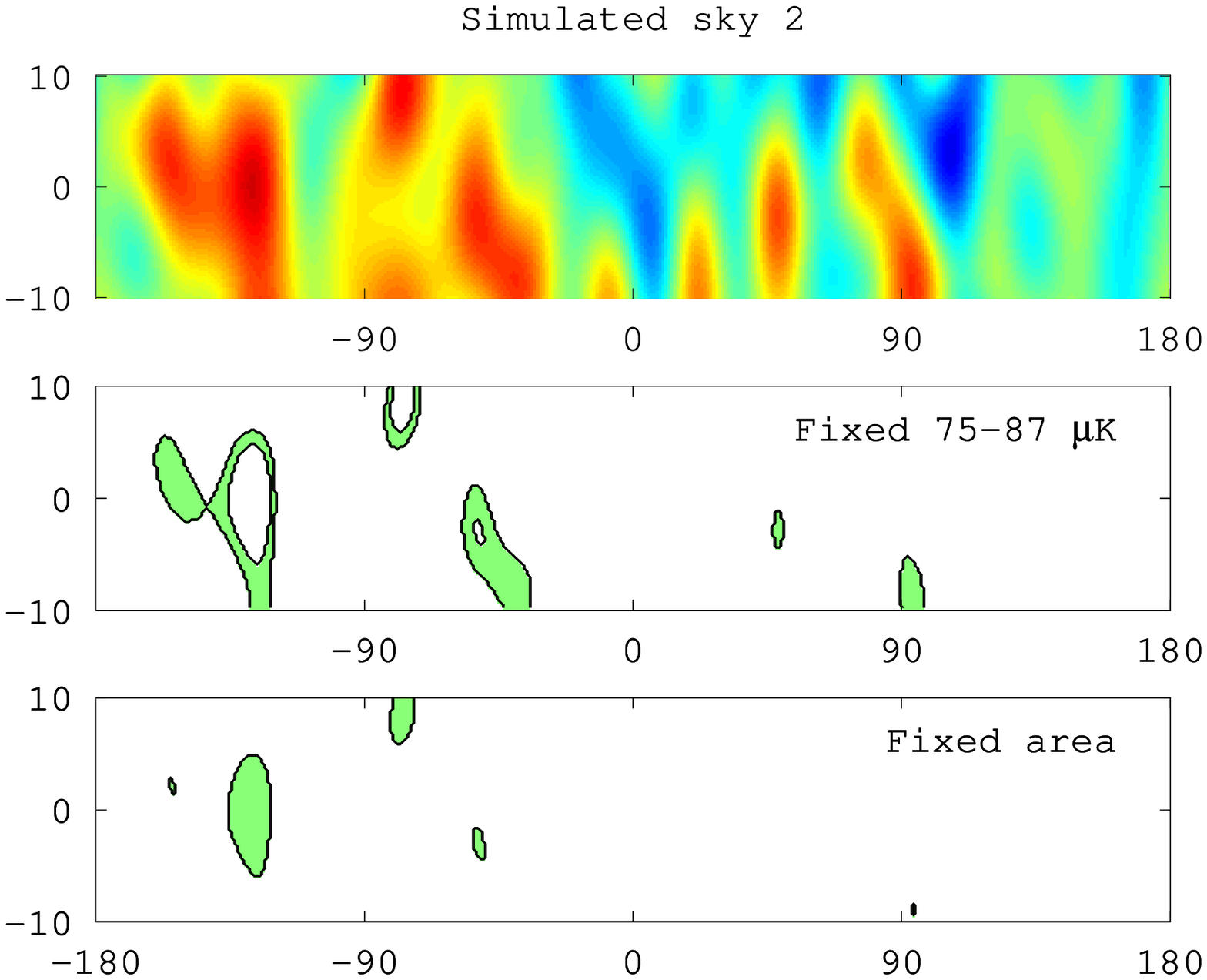}
   \end{tabular}
   \end{center}
   \caption[example]
   { \label{fig:strip_sim2}
      A second simulated sky realization within the Loop I strip.  The color scale in the temperature map
      {\it (top panel)} is [$-130$,$+130~\mu\mathrm{K}$].  This realization has $T_\mathrm{max}$
      higher than in the ILC map.  The temperature range $\lmsbinfourtmin\le T<\lmsbinonetmax~\mu\mathrm{K}$ {\it (middle panel)}
      does not contain the hottest pixels, but instead selects moderately warm regions that are extended in shape
      and cover a large solid angle.  About $\pctthousand\%$ of the
      simulated sky maps fall into this category.  The true peaks can be
      selected by taking the hottest pixels up to a fixed area {\it (bottom panel)}.
   }
   \end{figure}
%-------------

For each bin, a low value of $G_i$ indicates that the hottest pixels in the filtered WMAP ILC map
are clustered near the Loop location.  This is compared with the equivalent values $g_i$ in
the simulated sky realizations.  The $p$-value is calculated as the fraction of simulated
realizations for which $g_i < G_i$.  The $p$-values from $LMS$ are summarized in Tab.~\ref{tab:stats}.
The majority of the statistical power in the cluster analysis 
comes from the highest bin, $\lmsbinonetmin\le T<\lmsbinonetmax~\mu\mathrm{K}$, containing the
hottest parts of the hot spots.  This bin has clustering statistic $G_1=0.517^\circ$.

%%%%%%%%%%%%%%%%%%%%%%%%%%%%%%%%%%%%%%%%%
\section{Selection of hottest pixels}

I have compared the properties of the WMAP ILC map using $\rwonsim$ simulated sky realizations
generated with Synfast from the \healpix\ package~\citep{healpix}, then filtered to $\ell\le 20$ and
downgraded to NSIDE=128, as in $LMS$.  This same set of $\rwonsim$ simulations will be analyzed in
several different ways to reproduce and test the $LMS$ cluster analysis.
These maps show a wide variation in
the temperature of the hottest pixel, as in Fig.~\ref{fig:tmax_hist}.
The hottest single pixel in the annulus in the filtered WMAP ILC map has $T=\lmsmaxpix~\mu\mathrm{K}$.
Across simulated realizations, the mean value of the hottest pixel temperature is $\sim\tmaxmean~\mu\mathrm{K}$.
For $\tmaxpctabove\%$ of the simulated realizations, the hottest pixel temperature is higher than that
of the WMAP ILC map.  This does not indicate any inconsistency between the real sky and the
simulations, but simply shows that the peaks in different realizations are found
at different temperatures, typically $60$--$200~\mu\mathrm{K}$.

Although the peak temperatures are different in different realizations, the analysis in $LMS$ has used
fixed bins in the range $75$--$87~\mu\mathrm{K}$.  Because these bins have been chosen with reference
to the WMAP ILC map, they will contain the hot spots for this map.  However, the hottest pixels do not generally fall
into these bins for a simulated sky map.  I show two such mock maps in
Fig.~\ref{fig:strip_sim1} and Fig.~\ref{fig:strip_sim2}.  The first of these has a somewhat lower $T_\mathrm{max}$ than
the WMAP ILC map, and as a result has zero pixels within the four bins.  Of the 
$\rwonsim$ realizations calculated here, $\nzero$ ($\pctzero\%$) fall into this category, represented by the
example in Fig.~\ref{fig:strip_sim1}.

A much larger number of simulated maps -- most of the realizations -- have a higher-$T$
hottest pixel, as shown in the representative example in Fig.~\ref{fig:strip_sim2}.
These have a large number of pixels within the four used bins, but they do not include the peaks.
Instead, these bins now select a range of moderately warm spots that run around the peaks as shown in
the middle panel of Fig.~\ref{fig:strip_sim2}.  These regions are by nature extended rather than localized,
and their extended structure gives high values of the clustering statistic $g_i$.
A total of $\nthousand$ realizations ($\pctthousand\%$) fall into this category, defined as those realizations with
at least 100 pixels in the $84$--$87~\mu\mathrm{K}$ bin (compared to 33 pixels in the WMAP ILC map).

The use of fixed-temperature bins therefore represents an unfair tailoring of the analysis to the
WMAP ILC map, since the bins are guaranteed to identify the peaks in the ILC map but not
in the simulations.  A simulation whose hottest pixels do (by chance) cluster at the Loop location will typically
not have a low value of $g_i$ unless its hottest pixels also (again by chance) have a temperature
between 84 and 87~$\mu$K.
The comparison of $g_i$ from the simulations with $G_i$ from the ILC
as performed in $LMS$ therefore fails to give a correct estimate of the likelihood of
the observed clustering.  In the next section I propose a modified binning that selects
the peak pixels in each realization, and I calculate corrected $p$-values.

%%%%%%%%%%%%%%%%%%%%%%%%%%%%%%%%%%%%%%%%%
\section{Modified peak selection}

%-------------
   \begin{figure}[t]
   \begin{center}
   \begin{tabular}{c}
   \includegraphics[width=0.96\columnwidth]{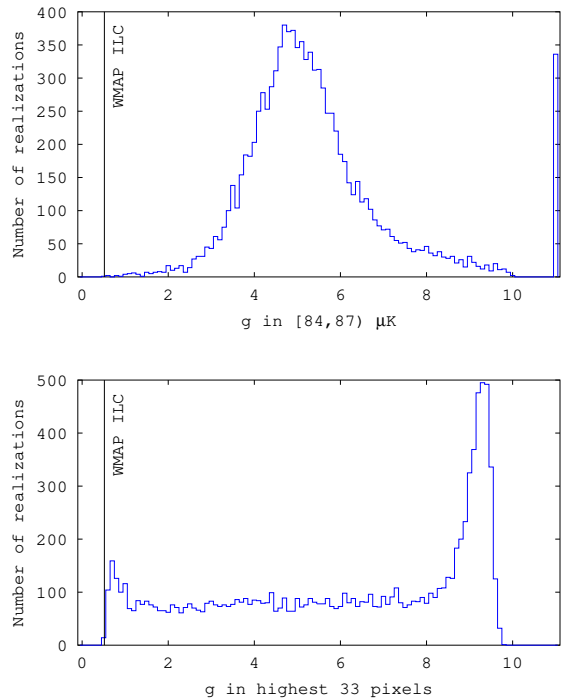}
   \end{tabular}
   \end{center}
   \caption[example]
   { \label{fig:gcmp_hist}
     \label{fig:gfix_hist}
     \label{fig:gtop_hist}
     Radial clustering statistic as calculated using fixed binning in $T$, as in $LMS$ {\it (top panel)},
     and using a fixed number of pixels {\it (bottom panel)}.  For most realizations, the fixed bins in
     $T$ contain pixels with moderately warm temperatures and not the true peaks.  These have extended
     patterns that often cross the width of the annulus and give $g_i=5$.  This is the reason for the
     central distribution in the top panel.  A small but significant number of realizations have zero pixels
     within the bin, and appear here at $g_i=11$.  When the peaks are properly selected in each realization,
     the distribution of $g_i$ becomes nearly flat.  There is a peak at the high end caused by realizations
     in which the brightest pixels are not local maxima, but sloping edges that intersect the inner or
     outer edge of the annulus.  The smaller peak at the low end corresponds to the minimum value of $g_i$
     set by the size of a single hot spot after filtering to $\ell\le 20$.
   }
   \end{figure}
%-------------

To calculate a correct statistic, one should select the pixels in the simulated
maps in a way that matches the treatment of the real data.  A simple and valid way to do this is to choose the hottest $N$ pixels, where
$N$ is taken from the number of pixels in the same bin in the ILC map.  This is equivalent to choosing
the same number of pixels in each realization, so that
$N\left(A_i(j^\mathrm{th}~\mathrm{sim})\right)=N\left(A_i(\mathrm{ILC})\right)$.
Using this modified definition, the clustering statistic $G_i$ for the WMAP ILC is unchanged,
but the statistics $g_i$ for the simulations are now different.

I have compared the results of cluster analysis on my $\rwonsim$ simulated sky realizations,
both with the unmodified 75--87~$\mu$K bins and with fixed pixel count in each bin.
The results are shown in Fig.~\ref{fig:gfix_hist}.
The top panel shows the distribution of $g_1$
simulations and the WMAP ILC map, using fixed $\lmsbinonetmin\le T<\lmsbinonetmax~\mu\mathrm{K}$
as in $LMS$.
The values are grouped around $5^\circ$, indicating
that the selected pixels are not preferentially located near the center line of the
strip or near the edges.  This reflects the fact that the selected regions are generally
not the hottest pixels, but sinuous intermediate contours as in the middle panel of
Fig.~\ref{fig:strip_sim1}.  The spike at the right side of the histogram
represents those realizations that have no pixels within the $T$-bin.

% %-------------
%    \begin{figure}
%    \begin{center}
%    \begin{tabular}{c}
%    \includegraphics[width=0.96\columnwidth]{gfix_hist}
%    \end{tabular}
%    \end{center}
%    \caption[example]
%    { \label{fig:gfix_hist}
%      Radial clustering statistic as calculated using fixed binning in $T$
%    }
%    \end{figure}
% %-------------

The results of the modified analysis are shown
in the bottom panel of Fig.~\ref{fig:gtop_hist}.  The $g_1$ values are
now distributed evenly across the interval $0<g_1<10^\circ$, with some pileup near $g_1=10$.
I have recalculated all the statistics from Table~1 of $LMS$ after
changing the treatment of the simulated maps to select the true hot spots as described above.  The results are
given in the middle column of Table~\ref{tab:stats}.
This calculation shows an overall significance that is weaker than reported in the $LMS$ article.
The test with all four bins has a recalculated $p$-value of $\sim 0.1\%$ instead of $10^{-4}$.

%-------------
\begin{table}[t]
\caption{Probability of $g_i<G_i$ ($i=1$--$4$) evaluated with $\rwonsim$ simulations}
\label{tab:stats}
\begin{center}
\begin{tabular}{lrrr}
\hline \hline
\rule[-1ex]{0pt}{3.5ex}  Criterion & Probability & Corrected & Corrected \\
\rule[-1ex]{0pt}{3.5ex}  & ($LMS$) & NSIDE=128 & NSIDE=512 \\
\hline
\rule[-1ex]{0pt}{3.5ex}  $g_i<G_i$ for all 4 bins & 0.018\% & $\sim0.1\%$ & 0.5\% \\
\rule[-1ex]{0pt}{3.5ex}  $g_i<G_i$ for any 3 in 4 bins & 2.0\% & 5.0\% & 5.8\% \\
\rule[-1ex]{0pt}{3.5ex}  $g_i<G_i$ for any 2 in 4 bins & 3.3\% & 13.9\% & 13.9\% \\
\rule[-1ex]{0pt}{3.5ex}  $g_i<G_i$, $i=1$ ($T=84~\mu\mathrm{K}$) & 0.1\%\footnote{When replicating the $LMS$ analysis without modification I obtain $\sim 0.01\%$ rather than $0.1\%$ as reported in Table~1 of $LMS$.} & $\sim0.1\%$ & 1.2\% \\
\rule[-1ex]{0pt}{3.5ex}  $g_i<G_i$, $i=2$ ($T=81~\mu\mathrm{K}$) & 2.9\% & 17.0\% & 17.9\% \\
\rule[-1ex]{0pt}{3.5ex}  $g_i<G_i$, $i=3$ ($T=78~\mu\mathrm{K}$) & 7.5\% & 15.4\% & 14.8\% \\
\rule[-1ex]{0pt}{3.5ex}  $g_i<G_i$, $i=4$ ($T=75~\mu\mathrm{K}$) & 8.6\% & 13.9\% & 14.7\% \\
\rule[-1ex]{0pt}{3.5ex}  $\Sigma_i g_i<\Sigma_i G_i$, $i=1$--$4$ & 1.0\% & 6.7\% & 7.4\% \\
\hline
\end{tabular}
\end{center}
\end{table}
%------------

%%%%%%%%%%%%%%%%%%%%%%%%%%%%%%%%%%%%%%%%%
\section{Dependence on map pixelization}

The clustering statistic $G_i$ is defined in terms of the locations of pixel centers, and the results will
therefore depend on the map pixelization used.  The analysis in $LMS$ uses \healpix\ maps at NSIDE=128, for
which the typical pixel size is $\sim 0.5^\circ$.  This is similar to the value of the clustering statistic
in the highest temperature bin, $G_1=0.517^\circ$.  For distances similar to or smaller than the
pixel size, the value of $G_i$ is highly sensitive to the locations of the pixel centers relative to the
circle defining Loop I.  A low value of $G_i \approx 0.5^\circ$ is therefore likely to depend on a coincidental arrangement
of \healpix\ pixels and not only on the properties of the WMAP ILC map.

In order to test this, I have repeated the same analysis at a finer map resolution, NSIDE=512, for which
the typical pixel dimension is $\sim 0.1^\circ$.  The results are shown
in the rightmost column of Table~1, continuing to apply modified binning as in Section~4.
When analyzed at finer map resolution, the clustering values
in the ILC map have $p$-values of $1.2\%$ (highest bin only) or $0.5\%$ (highest four bins) relative to the
simulations.  This shows that the high significance found by $LMS$ is partly attributable to details
of the pixelization scheme, and that the significance decreases when working at a map resolution high enough
to be insensitive to this effect. 

%%%%%%%%%%%%%%%%%%%%%%%%%%%%%%%%%%%%%%%%%
\section{Likelihood under Loop I interpretation}

The previous sections have shown that the reported significance is
reduced after correcting the selection of temperature peaks and using a finer map resolution to avoid pixelization
effects.  Even after these corrections, the clustering statistic still has a $p$-value $\sim 1\%$.  This remains
moderately unlikely under the assumption of statistical isotropy of the CMB as represented by the WMAP ILC
map.  However, such a result does not automatically constitute a detection of Loop I in the ILC map.
It is also necessary to show that a plausible model of contamination from Loop I can account for the
low $G_i$ values.

I have tested this using a simple model of Loop I power.
Power was added along a thin ring at the coordinates of Loop I, with
an amplitude chosen to match the $23.9~\mu\mathrm{K}$ mean temperature anomaly reported by $LMS$.
In an additional $\rwonsim$ simulations using modified binning (as in Section 4), at NSIDE=128,
I find $g_i<G_i$ in $\sim 1\%$ of the realizations.
This means that the observed value $G_i$ of the clustering statistic remains unlikely at the $\sim 1\%$ level even
under a model including power from Loop I in the ILC map.  This result could be modified under
a different model of Loop I power: for example, one with a variable brightness or anisotropic
distribution around the loop.  A likelihood ratio test can be used to test the compatibility
of the observed ILC map with a given Loop I model, accounting for any added degrees of freedom
in the model.

Such a test should also account for uncertainty in the precise position and shape of Loop I.
The position of Loop I is defined by the coordinates listed in~\cite{BHS}, which give a center of
$\left(l=329^\circ\pm 1.5^\circ\mathrm{,~}b=+17.5^\circ \pm 3^\circ\right)$ and a diameter of $116^\circ \pm 4^\circ$.
A clustering value of $G_1=0.517^\circ$ requires the peaks in the WMAP ILC map to line up
with the adopted Loop I center and radius to a precision much finer than the stated uncertainty
on these parameters, and even to a level smaller than the number of significant figures with which
these parameters have been given.  \cite{BHS} allow the Loop I features to deviate somewhat from
ideal circularity and to have finite thickness, stating an RMS deviation of $0.9^\circ$, which
is less than the value of $G_1$ in the $LMS$ cluster analysis.  These effects could be
taken into account in a likelihood ratio test by including the center position and radius of
Loop I as parameters of the model.

Although a fuller analysis should be performed, a test against a simple model suggests that
the observed clustering statistics may be unlikely even under the assumption that Loop I
contributes to the ILC map.  The uncertainty in the center, radius, and shape of Loop I also
indicate that the observed clustering must be attributed at least partly to chance under
any such hypothesis.

%%%%%%%%%%%%%%%%%%%%%%%%%%%%%%%%%%%%%%%%%
\section{Conclusions}

The significance of clustering evidence for Loop I contamination in the WMAP ILC is overestimated
in the $LMS$ article because the hottest pixels have not been correctly identified in
the simulated sky realizations, and because the coarse map resolution makes the
analysis sensitive to details of the pixelization scheme that are not physically meaningful.
Accounting for these gives a corrected $p$-value of $\sim 1\%$ rather than the $\sim 10^{-4}$
reported in $LMS$.  In addition, it should be demonstrated that the observed clustering
values are adequately explained by the interpretation that power from Loop I is present in
the WMAP ILC map.  In fact, I find that the clustering in the highest bin remains
unlikely, and must apparently be attributed to chance even under this interpretation.
The consistency of the ILC map with a hypothesis of Loop I power could be tested more
quantitatively by performing a likelihiood ratio test.
The additional analysis presented here does not rule out the existence at some level of contamination
from Loop I in the WMAP ILC map, but it does greatly reduce the statistical
significance attached to claimed evidence for such contamination.

% % % % % % % % % % % % % % % % % % % % % % % % % % % % % % %
% % % % % % % % % % % % % % % % % % % % % % % % % % % % % % %
\acknowledgements
I acknowledge support from a KIPAC Kavli Fellowship and the Department of Energy.
I am grateful to Chao-Lin Kuo and Sergi R. Hildebrandt for helpful discussion and comments.

\bibliographystyle{apj}
\bibliography{Loop}

\end{document}